\documentclass[12pt]{article}

\newcommand{\be}{\begin{equation}}
\newcommand{\ee}{\end{equation}}
\usepackage{epsfig,amsmath,amssymb,graphics,graphicx} 

\begin{document}

\begin{center}
{\it Journal of Physics: Condensed Matter 20 (2008) 175223}
\vskip 5 mm

{\Large \bf Universal Electromagnetic Waves in Dielectric}
\vskip 3 mm

{\large \bf Vasily E. Tarasov }

\vskip 3mm

{\it Skobeltsyn Institute of Nuclear Physics, \\
Moscow State University, Moscow 119991, Russia \\ 
E-mail: tarasov@theory.sinp.msu.ru}

\end{center}

\begin{abstract}
The dielectric susceptibility of a wide class of dielectric materials 
follows, over extended frequency ranges, 
a fractional power-law frequency dependence that is called
the "universal" response.
The electromagnetic fields in such dielectric media 
are described by fractional differential equations  
with time derivatives of non-integer order. 
An exact solution of the fractional equations for a magnetic field is derived.
The electromagnetic fields in the dielectric materials 
demonstrate fractional damping.
The typical features of "universal" electromagnetic waves 
in dielectric are common to a wide class of materials, 
regardless of the type of physical structure, 
chemical composition, or of the nature of the polarizing species, 
whether dipoles, electrons or ions.
\end{abstract}



\newpage
\section{Introduction} 

A growing number of dielectric relaxation data show that 
the classical Debye behavior \cite{W1} 
is hardly ever observed experimentally 
\cite{Jo2,Jrev,Ram}. 
The fact that different dielectric spectra are
described by the power laws is confirmed in
dielectric measurements realized by Jonscher \cite{Jo2,Jrev} 
for a wide class of various substances.
The dielectric susceptibility of most materials follows, 
over extended frequency ranges, a fractional power-law frequency dependence, 
which is called the law of "universal" response  \cite{Jo2,Jrev}. 
This law is found both in dipolar materials beyond 
their loss-peak frequency, and in materials where the polarization 
arises from movements of either ionic or electronic hopping charge carriers.
It has been found \cite{Jo4,Jo6} that 
the frequency dependence of the dielectric susceptibility
$\tilde \chi(\omega)=\chi^{\prime}(\omega)-i\chi^{\prime \prime}(\omega)$
follows a common universal pattern for virtually all kinds of materials. 
Namely, the behavior
\be \label{W-3}
\chi^{\prime}(\omega) \sim \omega^{n-1} , \quad 
\chi^{\prime \prime}(\omega) \sim \omega^{n-1} , \quad 
(0 < n < 1, \quad \omega \gg \omega_p) ,
\ee
and
\be  \label{W-4}
\chi^{\prime}(0)-\chi^{\prime}(\omega) \sim \omega^{m} , \quad 
\chi^{\prime \prime}(\omega) \sim \omega^{m} , \quad 
(0 < m < 1, \quad \omega \ll \omega_p) ,
\ee
where $\chi^{\prime}(0)$ is the static polarization 
and $\omega_p$ the loss-peak frequency, 
is observed over many decades of frequency. 
Note that the ratio of the imaginary to the real component
of the susceptibility is independent of frequency. 
The frequency dependence given by equation (\ref{W-3}) 
implies that the real and
imaginary components of the complex susceptibility obey 
at high frequencies the relation 
\be \label{W-5}
\frac{\chi^{\prime \prime}(\omega)}{\chi^{\prime }(\omega)} =
\cot \left( \frac{\pi n}{2} \right) , \quad  (\omega \gg \omega_p) .
\ee
The experimental behavior (\ref{W-4}) 
leads to a similar frequency-independent rule 
for the low-frequency polarization decrement:
\be \label{W-6}
\frac{\chi^{\prime \prime}(\omega)}{ \chi^{\prime }(0)- \chi^{\prime }(\omega) }
=\tan \left( \frac{\pi m}{2} \right) , \quad (\omega \ll \omega_p) .
\ee

There are many models that describe and explain 
the universal response laws. Let us note some of them. 
A general equation for the susceptibility of disordered systems 
is proposed in \cite{Bergman}. 
It is shown  \cite{Bergman} that this equation 
is a good frequency domain representation of 
the time domain Kohlrausch-Williams-Watts stretched exponential. 
In \cite{BSC}, the authors have derived a general 
two-power-law relaxation function for heterogeneous materials
using the maximum entropy principle for nonextensive systems. 
The asymptotic power-law behaviors 
coincide with those of the Weron 
generalized dielectric function derived in the stochastic theory 
from an extension of the Le'vy central limit theorem. 
These results \cite{BSC} are in agreement with the 
Jonscher universality principle. 
The memory function approach and scaling relationships 
can be used \cite{RF2} as a basis for the dynamic model of 
symmetric dielectric spectrum broadening. 
For the correspondence between
the relaxation time, the geometrical properties, the self-diffusion 
coefficient and the exponent of
power-law wings was established in \cite{RF2}. 
In \cite{RF1},  a memory function equation and scaling relationships 
were used for the physical interpretation of the Cole-Cole exponent. 
The correspondence between the relaxation
time, the geometrical properties, the self-diffusion coefficient and 
the Cole-Cole exponent was considered in \cite{RF1}. 
The fractional power laws frequently observed in dielectric measurements 
can be interpreted \cite{Frenning} in terms of regular singular points of 
the underlying rate equation for the process giving rise to the dielectric response. 
The review \cite{Jrev} presents a wide-ranging broad-brush picture 
of dielectric relaxation in solids, making use of 
the existence of a "universality" of dielectric response regardless 
of a wide diversity of materials and structures, 
with dipolar as well as charge-carrier polarization.

The theory of integrals and derivatives of non-integer order goes back 
to Leibniz, Liouville, Riemann, Grunwald, and Letnikov \cite{SKM,KST}. 
Fractional analysis has found many
applications in recent studies in mechanics and physics.
The interest in fractional equations has been growing continually 
during the last few years because of numerous applications. 
In a short period of time the list of applications has become long
(see for example \cite{Zaslavsky1,Zaslavsky2,Mainardi,MK,Hilfer,Sing}). 
Note also the application of fractional calculus to the problems of
classical electrodynamics \cite{E1,Plasma2005}. 

In \cite{NR} fractional calculus has been undertaken 
in order to understand the nature of one type of nonexponential relaxation.
An equation containing operators of fractional integration and
differentiation is obtained and solved \cite{NR}, 
which the relaxation function obeys in this case. 
In \cite{NP}, 
the potential of the fractional derivative technique is demonstrated  
using the example of the derivation of all three known patterns of anomalous, 
nonexponential dielectric relaxation of an inhomogeneous medium 
in the time domain. 
It is explicitly assumed \cite{NP} that the fractional derivative 
is related to the dimensionality of a temporal fractal ensemble 
in a sense that the relaxation times are distributed 
over a self-similar fractal system. 
In \cite{Dielectric}, the model of dielectric relaxation based on 
the fractional kinetics containing the complex power-law exponents was used 
to received a confirmation in description of the real part of 
the complex conductivity by the fitting function.
The coincidence between the suggested model and measured data 
gives \cite{Dielectric} the possibility to suggest a more reliable 
physical picture of the swelling process that takes place 
in neutral and charged gels.
The real process of dielectric relaxation during the polymerization 
reaction can be described \cite{First} in terms of 
the fractional kinetic equations containing complex-power-law exponents. 
Based on the physical and geometrical meanings of the fractional integral 
with complex exponents there is a possibility \cite{First} of developing a model 
of dielectric relaxation based on the self-similar fractal character of 
the averaged microprocesses that take place in the mesoscale region. 

In this paper, we prove that in a time domain the 
fractional power-law frequency dependence 
gives differential equations 
with derivatives and integrals of non-integer order.
We obtain equations that describe "universal" 
electromagnetic waves for such dielectric materials.
The power laws are presented by fractional differential equations 
such that the electromagnetic fields in the materials 
demonstrate "universal" fractional damping.
An exact solution of the fractional equations for magnetic field is derived.
The electromagnetic fields in the dielectric materials 
demonstrate fractional damping.
The suggested fractional equations are common (universal)
to a wide class of materials, regardless of the type 
of physical structure, chemical composition or 
of the nature of the polarizing species, 
whether dipoles, electrons or ions.
A possible link between dielectric response
and structure of low-loss dielectrics is discussed.


\section{Fractional equations for universal laws}

For the region $\omega \gg \omega_p$, 
the universal fractional power law (\ref{W-3})
can be presented in the form
\be \label{chi-1}
\tilde \chi(\omega)= \chi_{\alpha} \, (i \omega)^{-\alpha} , 
\quad (0<\alpha<1) 
\ee
with some positive constant $\chi_{\alpha}$ and $\alpha=1-n$. 
Here
\[ (i\omega)^{\alpha}=|\omega|^{\alpha} \, e^{i \, \alpha \, \pi \, sgn(\omega)/2}. \]
It is easy to see that relation (\ref{W-5}) is satisfied.

The polarization density can be written as
\[ {\bf P}(t,r)={\cal F}^{-1} \left( \tilde {\bf P}(\omega,r) \right)=
\varepsilon_0 {\cal F}^{-1} \left(\tilde \chi(\omega)  
\tilde {\bf E}(\omega,r) \right) =
\varepsilon_0 \chi_{\alpha} {\cal F}^{-1} 
\left( (i\omega)^{-\alpha} \tilde {\bf E}(\omega,r) \right) , \]
where $\tilde {\bf P}(\omega,r)$ is a Fourier transform ${\cal F}$ of ${\bf P}(t,r)$. 

Note that the Fourier transform ${\cal F}$ 
of the fractional Liouville integral \cite{SKM,KST}
\[ (I^{\alpha}_{+}f)(t)=\frac{1}{\Gamma(\alpha)} 
\int^{t}_{-\infty} \frac{f(t') dt'}{(t-t')^{1-\alpha}}  \]
is given by the following result 
(see Theorem 7.1 in \cite{SKM} and  Theorem 2.15 in \cite{KST}):
\[ ({\cal F} I^{\alpha}_{+}f)(\omega)=
\frac{1}{(i\omega)^{\alpha}} ({\cal F}f)(\omega) . \]

Using the fractional Liouville integral, 
the fractional power-law (\ref{chi-1}) for $\tilde\chi(\omega)$ 
in the frequency domain gives
\be \label{Alpha}
{\bf P}(t,r)=\varepsilon_0 \chi_{\alpha} \, (I^{\alpha}_{+} {\bf E})(t,r) ,  
\quad (0<\alpha<1) . 
\ee
This equation means that polarization density ${\bf P}(t,r)$
for the high-frequency region is proportional to 
the fractional Liouville integral of the electric field.


For the region $\omega \ll \omega_p$, 
the universal fractional power law (\ref{W-4}) can be presented as
\be \label{chi-2}
\tilde \chi(\omega)=\tilde \chi(0)-
\chi_{\beta} (i \omega)^{\beta} ,  \quad (0<\beta<1) 
\ee
with some positive constants $\chi_{\beta}$, 
$\tilde \chi(0)$, and $\beta=m$. 
It is not hard to prove that equation (\ref{W-6}) is satisfied.

Note that the Fourier transforms ${\cal F}$ of
the fractional Liouville derivative \cite{SKM,KST} 
\[ (D^{\beta}_{+}f)(t)=\frac{\partial^k}{\partial t^k}(I^{k-\beta}_{+}f)(t)
=\frac{1}{\Gamma(k-\beta)} \frac{\partial^k}{\partial t^k}
\int^{t}_{-\infty} \frac{f(t') dt'}{(t-t')^{\beta-k+1}} , 
\quad (k-1 < \beta <k) \]
are given by the following result 
(see Theorem 7.1 in \cite{SKM} and Theorem 2.15 in \cite{KST}):
\[ ({\cal F} D^{\beta}_{+}f)(\omega)=(i\omega)^{\beta} \, ({\cal F}f)(\omega) . \]

Using the definition of the fractional Liouville derivative,  
the fractional power law (\ref{chi-2}),  gives the polarization density 
\[ {\bf P}(t,r)= 
\varepsilon_0 \, {\cal F}^{-1} 
\left(\tilde \chi(\omega)  \tilde {\bf E}(\omega,r) \right) \]
in the form
\be \label{Beta}
{\bf P}(t,r)= \varepsilon_0 \tilde \chi (0) \, {\bf E}(t,r) 
- \varepsilon_0 \chi_{\beta} \, (D^{\beta}_{+} {\bf E})(t,r) ,  
\quad (0<\beta<1) . \ee
This equation means that polarization density ${\bf P}(t,r)$
for the low-frequency region is defined by 
the fractional Liouville derivative of the electric field.

Equations (\ref{Alpha}) and (\ref{Beta}) can be considered as the
universal laws in the time domain. These equations allow us to derive
fractional wave equations for electric and magnetic fields.


\section{Universal electromagnetic wave equation}

Using the Maxwell equations, it is easy to obtain the equation
\be \label{5}
\varepsilon_0 \frac{\partial^2 {\bf E}(t,r)}{\partial t^2}+
\frac{\partial^2 {\bf P}(t,r)}{\partial t^2}+
\frac{1}{\mu } \left( grad \, div {\bf E}- \nabla^2 {\bf E} \right) +
\frac{\partial {\bf j}(t,r)}{\partial t}=0 . 
\ee

For the region $\omega \gg \omega_p$, 
the polarization density ${\bf P}(t,r)$ is related with ${\bf E}(t,r)$
by equation (\ref{Alpha}). 
Substituting (\ref{Alpha}) into (\ref{5}), 
we obtain the fractional equation for electric field
\be \label{EFE1}
\frac{1}{v^2} \frac{\partial^2 {\bf E}(t,r)}{\partial t^2}+
\frac{\chi_{\alpha}}{v^2} (D^{2-\alpha}_{+} {\bf E})(t,r) +
\left( grad \, div {\bf E}- \nabla^2 {\bf E} \right) =
-\mu\frac{\partial {\bf j}(t,r)}{\partial t} , \quad (0<\alpha<1) ,
\ee
where $v^2=1/(\varepsilon_0 \mu)$.
Note that $div \, {\bf E}\ne 0$ for $\rho(t,r)=0$.

For the region $\omega \ll \omega_p$, 
the fields ${\bf P}(t,r)$ and ${\bf E}(t,r)$ are connected
by equation (\ref{Beta}). Then equation (\ref{5}) gives
\be \label{EFE2}
\frac{1}{v^2_{\beta}} \frac{\partial^2 {\bf E}}{\partial t^2}
-  \frac{a_{\beta}}{v^2_{\beta}}  (D^{2+\beta}_{+} {\bf E})+ 
\left( grad \, div {\bf E}- \nabla^2 {\bf E} \right) =
-\mu\frac{\partial {\bf j}}{\partial t} , \quad (0<\beta<1) ,
\ee
where
\[ v^{2}_{\beta}=\frac{1}{ \varepsilon_0 \mu \, [1+  \tilde \chi (0)] } , \quad
a_{\beta}= \frac{\chi_{\beta}}{1+  \tilde \chi (0) } . \]
Equations (\ref{EFE1}) and (\ref{EFE2}) describe the time-evolution of 
the electric field in the dielectric materials.
These equations are fractional differential equations.


Using the Maxwell equations, it is not hard to obtain 
the equation for a magnetic field
\be  \label{mag5}
\frac{\partial^2 {\bf B}(t,r)}{\partial t^2}= 
\frac{1}{\varepsilon_0  \mu} \nabla^2 {\bf B}(t,r)+
\frac{1}{\varepsilon_0} \frac{\partial  }{\partial t} curl {\bf P}(t,r)+
\frac{1}{\varepsilon_0} curl {\bf j}(t,r). 
\ee

The experimental applied field ${\bf B}(t,r)$ can be presented as 
\[  {\bf B}(t,r)=
\begin{cases} 
0 , & t \le 0 ,
\cr 
{\bf B} (t,r) , & t > 0 .
\end{cases}
\]
For the region $\omega \gg \omega_p$, 
the polarization density ${\bf P}(t,r)$ is related to ${\bf E}(t,r)$
by equation (\ref{Alpha}), and
we obtain the fractional equation for magnetic field
\be \label{MFE1}
\frac{1}{v^2}\frac{\partial^2 {\bf B}(t,r)}{\partial t^2}+
\frac{\chi_{\alpha}}{v^2} \left( _0D^{2-\alpha}_{t} {\bf B} \right)(t,r)
- \nabla^2 {\bf B}(t,r) =\mu \, curl \, {\bf j}(t,r) ,  \quad (0<\alpha<1) ,
\ee
where $v^2=1/(\varepsilon_0  \mu)$, and $_0D^{2-\alpha}_{+}$
is the Riemann-Liouville derivative \cite{KST} on $[0,\infty)$ such that
\[ ( _0D^{2-\alpha}_{+}f)(t)=
\frac{1}{\Gamma(\alpha)} \frac{\partial^2}{\partial t^2}
\int^{t}_{0} \frac{f(t') dt'}{(t-t')^{1-\alpha}} , 
\quad (0 < \alpha <1) . \]

For the region $\omega \ll \omega_p$, we obtain
\be  \label{MFE2}
\frac{1}{v^2_{\beta}} \frac{\partial^2 {\bf B}(t,r)}{\partial t^2}-
\frac{a_{\beta}}{v^2_{\beta}} \left( _0D^{2+\beta}_{t} {\bf B} \right)(t,r)
- \nabla^2 {\bf B}(t,r) =\mu \, curl \, {\bf j}(t,r) , \quad (0<\beta<1) ,
\ee
where
\[ v^2_{\beta} =\frac{1}{\varepsilon_0 \mu \, [1+ \tilde \chi (0) ]} , \quad
a_{\beta}=\frac{\chi_{\beta} }{1+ \tilde \chi (0)} . \]
Equation (\ref{MFE2}) is a fractional differential equation that 
describes magnetic field in dielectric media.

\section{Fractional damping of magnetic field}

Let us obtain a solution of equations (\ref{MFE1}) and (\ref{MFE2}).
These equations can be represented in a general form. 
This general form of the fractional equations for magnetic field is
\be \label{MM}
( _aD^{\alpha}_t {\bf B})(t,r)-
\lambda_1  \, \left( _aD^{\beta}_{t} {\bf B} \right)(t,r)
- \lambda_2 \, \nabla^2 {\bf B}(t,r) ={\bf f}(t,r), 
\quad (1\le\beta<\alpha<3) ,
\ee
where ${\bf B}(a,r)=0$,
and the curl of current density of free charges 
is considered as an external source term:
\[ {\bf f}(t,r)= \mu \lambda_2 \, curl \, {\bf j}(t,r) . \]
Equation (\ref{MM}) gives equation (\ref{MFE1}) for $\alpha=2$, $1<\beta<2$, and
\[ \lambda_1= - \chi_{\alpha}, \quad \lambda_2=v^2=1/(\varepsilon_0  \mu) . \]
Equation (\ref{MFE2}) can be derived from (\ref{MM}) 
with $2<\alpha<3$, $\beta=2$, and
\[ \lambda_1= \frac{1}{a_{\beta}}=
\frac{1+\tilde \chi(0)}{\chi_{\beta}} , \quad 
\lambda_2=-\frac{v^2_{\beta}}{a_{\beta}}=
\frac{-1}{\varepsilon_0\mu \chi_{\beta}} . \]

An exact solution of (\ref{MM}) can be presented in terms of 
the Wright functions \cite{KST,MGo}.
Using the three-dimensional Fourier transform of equation (\ref{MM})
with respect to coordinates and Theorem 5.5 in \cite{KST}, 
we obtain the solution 
\be \label{Btr}
{\bf B} (t,r)= \frac{1}{(2 \pi)^3}  \int_{\mathbb{R}^3} d^3 k
\int^{t}_0 \, dt' \ e^{i kr} \, \tilde {\bf f}(t',k) \, G(t-t',k) , \ee
where
\be 
G(t-t',k)=\sum^{\infty}_{s=0} \frac{(-k^2 \lambda_2)^s}{s!} \,
(t-t')^{\alpha s+\alpha-1} \
_1\Psi_1 \left[ \frac{(s+1,1)}{(\alpha s +\alpha,\alpha-\beta)} \Bigl| 
 \lambda_1 (t-t')^{\alpha-\beta} \right] .
\ee
Here, $ _1\Psi_1$ is the Wright functions
\[ 
_1\Psi_1 \left[ \frac{(s+1,1)}{(\alpha s +\beta,\alpha)} \Bigl| z \right] \, = \,
\sum^{\infty}_{j=0} 
\frac{\Gamma(s+j+1)}{\Gamma(\alpha s+\beta+\alpha j)} \frac{z^j}{j!} 
\]
that can be derived as $\partial^s E_{\alpha,\beta}[z] / \partial z^s$
of the Mittag-Leffler function  $E_{\alpha,\beta}[z]$ (see \cite{KST,MGo}).
Equation (\ref{Btr}) describes the fractional field damping of 
magnetic field in the dielectric media.
An important property of the evolution described by the fractional 
equations for electromagnetic waves 
is that the solutions have power-like tails.


\section{Possible link between dielectric response and molecular structure of dielectric}

An interesting problem of considerable practical importance 
is the link between the dielectric loss of low-loss materials
and the identifiable physical and structural features in the material.
The following questions relating to this subject can be formulated. 
What material conditions must be satisfied for the low-loss property to be observed?
What are the physical interpretations of this type of behavior?

Jonscher has suggested in \cite{Jo2,Scr1,Scr2}
that the key to the understanding of low-loss
behavior lies in dipolar screening which leads
to interactions between dipoles in materials with sufficiently low dipole densities. 
In other words, the Jonscher's proposition
is that electrostatic interactions between dipoles
even at large distances from one another lock these
dipoles so that their response to external alternating fields is weakened.
The physical basis of low-loss dielectrics 
can be connected \cite{Jo2,Scr1,Scr2}
with the effect of dipolar screening which
produces interaction between neighbouring dipoles,
thereby preventing a significant part of them from
following external alternating fields. 
It is evident that the case of complete screening must correspond to Debye relaxation, 
because there is no interaction between any of the dipoles. 
The result of dipole-dipole interactions
with incomplete screening extending over many
dipoles is that the system becomes insensitive in the sense
that the fields generated by the interacting dipoles are
stronger than externally applied fields \cite{Scr2}.

We can also assume that the effect of screening 
is very important for dipolar relaxation.
In \cite{Scr2}, the Debye screening of the static fields is considered,
but the form of dipole-dipole interaction is not discussed.
In this section, we consider a possible form of screening
dipole-dipole interaction for low-loss dielectric materials.
In general, the dipole feels an electric field both from the permanent
charges of the system and from the other induced dipoles. 
To describe low-loss dielectrics, we suggest using 
a modification of the molecular polarization method.
This polarizable point dipoles method has
been applied to a wide variety of atomic and molecular systems, ranging from
noble gases and water to amorphous materials and to proteins. 
Let us note the main features of the polarization methods
and a possible generalization to describe low-loss dielectrics. 
In the point dipoles method, a polarizability 
$\alpha$ is associated to one or more sites \cite{Applequist}.
This method for treating polarizability is to add point inducible dipoles
on some or all sites.

The total electric field acting on each site is produced by 
the external charges and by induced dipole moments 
\cite{Mazur,Ramshaw,Applequist,Rick,41,Masia2}, such that
\be \label{i1}
{\bf E}_n={\bf E}^0_n-\sum_{m \ne n} {\bf T}_{nm} {\bf p}_m ,
\ee
where ${\bf p}_m$ is a dipole moment, 
${\bf T}_{nm}$ denotes the dipole field tensor, and
${\bf E}^0_n$ is the static field at site $n$ due to the permanent charges
\be \label{i2}
{\bf E}^0_n = \sum_{m\ne n} \frac{q_m {\bf r}_{nm}}{r^3_{nm}} ,
\ee
where $q_m$ is the charge at site $m$ and
$r_{nm}=|{\bf r}_{nm}|$ is the distance between $n$ and $m$. 
The induced dipoles interact through the dipole field tensor
\be \label{i3} 
{\bf T}_{nm}=\frac{1}{r^3_{nm}} {\bf I}-\frac{3}{r^5_{nm}} {\bf r} \otimes {\bf r} , \ee
where ${\bf I}$ is the unit tensor (identity matrix).
Here ${\bf r} \otimes {\bf r}$ is the tensor with elements $x_k x_l$,
where $x_k$ $(k,l=1,2,3)$ are the Cartesian components of the vector ${\bf r}_{nm}$
between $n$ and $m$.
The dipole moment, ${\bf p}_n$, induced on a site $n$ is
proportional to the electric field ${\bf E}_n$ at that site, 
\[ {\bf p}_n= \alpha_n {\bf E}_n .  \]
The proportionality constant is the polarizability tensor $\alpha_n$. 
The induced point dipole on site $n$ 
can be computed iteratively until a given threshold of
convergence for the induced dipole is reached 
(for example, see \cite{Rick} for issues concerning the efficiency 
of the different methods for liquid state simulations).
The energy  of dipole-induced dipole interaction has the form
\[ U_{pp} =\frac{1}{2} \sum_{m} \sum_{m \ne n} {\bf p}_n {\bf T}_{nm} {\bf p}_m . \]

Here we consider that all the dipoles will interact through the dipole field tensor. 
The method of Applequist, Carl, and Fung, \cite{Applequist,40} 
for calculating molecular polarizabilities uses this approach. 
One problem with coupling all the dipoles
with the interaction given by equation (\ref{i3}) is the "polarization catastrophe". 
As pointed out by Applequist {\it et al} \cite{Applequist} and Thole \cite{41}, 
the polarization ${\bf p}_n$, and therefore the induced dipole moment, 
may become infinite at small distances. 
The polarization catastrophe is avoided by screening
(attenuating) the dipole-dipole interaction \cite{41}.  
As with the screening of the static field, screening of 
the dipole-dipole interaction can be physically interpreted 
as correcting for the fact that the electronic distribution
is not well represented by point charges and dipoles \cite{Applequist,41,43}. 
The Thole approach for screening is to introduce the screening (damping) functions.

Equations (\ref{i1}) - (\ref{i3}) retain their validity \cite{Masia2} 
with the only change being that both the electric field created by 
a fixed charge and that
created by a point dipole (depending on the molecular model) are 
screening by functions $f_1(r)$ and $f_2(r)$,
\be \label{i4}
{\bf E}^0_n = \sum_{m\ne n}  f_1(r_{nm})  \frac{q_m {\bf r}_{nm}}{r^3_{nm}}
\ee
and
\be \label{i5} 
{\bf T}_{nm}=f_1(r_{nm}) \frac{1}{r^3_{nm}} {\bf I} - 
f_2(r_{nm})\frac{3}{r^5_{nm}} {\bf r} \otimes {\bf r} . \ee
In the limit of point charges and point dipoles, we have $f_1(r)=f_2(r)=1$, 
and the usual expressions (\ref{i2}) and (\ref{i3}) are obtained.
If, on the contrary, they are thought to be spatially extended,
the form of the screening depends on the charge distribution assumed. 
Thole concluded that a linear decrease of charge
density (up to a cutoff $a$) was ideal for the purpose 
of fitting of the molecular polarizability for which it had been designed. 
In this approximation we have for the screening functions
\be \label{ff1}
f_1(r)=4 \left|\frac{r}{a}\right|^3- 3\left|\frac{r}{a}\right|^4 ,
\quad f_2(r)=\left|\frac{r}{a}\right|^4 .
\ee
The possible alternative is an exponential distribution \cite{G1,G2,Masia2}.
Note that various computer simulations with different screening functions have been used.

We assume that Thole linear decreasing should be replaced
by fractional power-law dependence for low-loss dielectrics.
Therefore we suggest that dipolar screening 
in low-loss dielectric materials can be described 
by the screening functions
\be \label{ff2}
f_1(r)=\alpha_2\left|\frac{r}{a}\right|^{\alpha_1} - 
\alpha_1 \left|\frac{r}{a}\right|^{\alpha_2} ,
\quad f_2(r)=\left|\frac{r}{a}\right|^{\alpha_2} ,
\ee
where $\alpha_1$ and $\alpha_2$ are non-integer positive numbers
that define the fractional parameters $\alpha=1-n$ and $\beta=m$  
of laws (\ref{W-3}) and (\ref{W-4}).
The simplest possible dependences are
\be \label{aa}
\alpha_1=3-\alpha, \quad \alpha_2=4+\beta . \ee
For $\alpha_1=3$ and $\alpha_2=4$, equations (\ref{ff2}) 
give the Thole screening functions (\ref{ff1}). 
Using functions (\ref{ff2}), we obtain
the field from the permanent charges in the form
\be \label{i2b}
{\bf E}^0_n = \sum_{m\ne n} \frac{q_m {\bf r}_{nm}}{ a^2 r_{nm}} \,
\Bigl\{ \alpha_2 \left| \frac{r_{nm}}{a} \right|^{\alpha_1-2}
- \alpha_1 \left| \frac{r_{nm}}{a} \right|^{\alpha_2-2} \Bigr\} ,
\ee
and the induced dipoles interact through the dipole field tensor
\be \label{i5b}  {\bf T}_{nm}=\sum_{m\ne n} 
\Bigl\{ \frac{\alpha_2}{a^3} \left| \frac{r_{nm}}{a} \right|^{\alpha_1-3} {\bf I} 
- \frac{ \alpha_1 r^2_{nm} {\bf I} + 3{\bf r} \otimes {\bf r}}{a^3 r^2_{nm}}
\left| \frac{r_{nm}}{a} \right|^{\alpha_2-3} \Bigr\} . \ee
For $\alpha_1=3$ and $\alpha_2=4$, equations (\ref{i2b}) and (\ref{i5b})
give the usual Thole screening. 
For treating polarizability of low-loss dielectric, 
we suggest use the fractional values of the parameters $\alpha_1$ and $\alpha_2$.
A possible generalization of exponential screening is discussed in the Appendix.

Our suggestion to use fractional power laws in (\ref{ff2}) 
is based on the mathematical results that 
have been obtained in \cite{JPA}. 
In this paper, we consider the vectors ${\bf r}_n$ that define sites
as in solid state physics \cite{Kittel}, i.e.,
\[ {\bf r}_{\bf n}=\sum^3_{i=1} n_i {\bf a}_i , \]
where ${\bf a}_i$ are the translational vectors of the lattice, then 
\[ {\bf r}_{nm}={\bf r}_{n}-{\bf r}_{m}=\sum^3_{i} (n_i-m_i) {\bf a}_i . \]
Note that this representation can be used as an approximation, and, 
in general, should be generalized for amorphous materials.
As a result, equations (\ref{i2b}) and (\ref{i5b}) describe 
power-law long-range interactions of the type
$|n-m|^{-s-1}$ with fractional values of $s$. 
In the continuous limit,  these interactions 
are described by fractional integro-differentiation \cite{JPA}.
Using the results of \cite{JPA}, we can state that 
dipole-dipole interactions with fractional screening 
are connected with integro-differentiation of non-integer order.

As a result, we assume that the link between the dielectric loss of low-loss materials
and the physical and structural features in the material
can be realized by the generalized polarizable point dipoles method 
with fractional screening 
in molecular dynamics and Monte Carlo simulations.
Using these simulations the presence of identified defects
in the structure of low-loss materials can be also taken into account.
To realize this suggested approach additional investigations are required.


\section{Conclusion}

In conclusion, the electromagnetic fields and waves in a wide class of 
dielectric materials must be described by 
fractional differential equations with derivatives of order 
$2-\alpha$ and $2+\beta$, where $0<\alpha<1$ and $0<\beta<1$.
The parameters $\alpha=1-n$ and $\beta=m$ are defined by exponents 
$n$ and $m$ of the "universal" response laws  
for frequency dependence of the dielectric susceptibility. 
A remarkable property of the dynamics described by the fractional 
equations for electromagnetic fields
is that the solutions have power-like tails.
The typical features of "universal" electromagnetic waves 
in dielectric are common to a wide class of materials, 
regardless of the type of physical structure, chemical composition
or of the nature of the polarizing species, whether dipoles, electrons or ions.
We assume that the link between the dielectric loss of low-loss materials
and the physical/structural features in the material
can be obtained by the polarizable point dipoles method 
with fractional screening of dipole-dipole interactions 
in the molecular dynamics and Monte Carlo simulations.

For small fractionality of $\alpha$ (or $\beta$), it is possible to use
$\varepsilon$-expansion \cite{TZ2} 
over the small parameter $\varepsilon=\alpha$ (or $\varepsilon=1-\beta$). 
There are several numerical methods 
to solve fractional equations (see for example \cite{Numer}).
Note that the suggested fractional differential equations,
which describe the electromagnetic field
in dielectric media with power-law 
response, can be solved numerically.
For example, the Grunwald-Letnikov discretization 
scheme \cite{SKM} can be used to compute fractional 
equations for electromagnetic field in dielectric.

\newpage
\section*{Appendix}

In section 5, we consider the Thole screening functions and
their fractional generalizations.
The alternative distributions that are used in the polarizable point dipole
method are the  exponential screening functions 
\be \label{ef0}
f_1(r)=1-\exp \Bigl\{ - \Bigl(\frac{r}{a} \Bigr)^3 \Bigr\}
\ee
\be
f_2(r)=1-\Bigl[1+ \Bigl(\frac{r}{a} \Bigr)^3 \Bigr]\, 
\exp \Bigl\{ - \Bigl(\frac{r}{a} \Bigr)^3 \Bigr\} .
\ee
If we consider the modification of $f_1(r)$ of the form
\be \label{ef1}
f_1(r)=
\left(1-\exp \Bigl\{ - b \Bigl(\frac{r}{a} \Bigr)^3 \Bigr\} \right) \,
\exp \Bigl\{ -  c \frac{r}{a}  \Bigr\} ,
\ee
then for $r \ll a$ and $b=4$, $c=3/4$,
we obtain the Thole function for the electric field created by fixed charge.
For $b=1$ and $c=0$, equation (\ref{ef1}) gives (\ref{ef0}).
For $r \gg a$, we can use the approximation
\be
f_1(r) = \exp \Bigl\{ - \frac{3}{4} \frac{r}{a}  \Bigr\} 
\ee
that is the Debye screening function with $\lambda_D=4a/3$ used  
by Jonscher in \cite{Jo2,Scr1,Scr2}.

Note that the suggested fractional screening function $f_1(r)$ 
of the form (\ref{ff2}) can be an approximation of the screening function 
\be \label{ef2}
f_1(r)=
\left(1-\exp \Bigl\{ - b \Bigl(\frac{r}{a} \Bigr)^{\alpha_1} \Bigr\} \right) \,
\exp \Bigl\{ -  \frac{\alpha_1}{\alpha_2} \Bigl(\frac{r}{a}\Bigr)^{\alpha_2-\alpha_1}  \Bigr\} ,
\ee
If $\alpha_1=3$ and $\alpha_2=4$, then equation (\ref{ef2}) gives (\ref{ef1}).
For $b=\alpha_2$ and $c=\alpha_1/\alpha_2$ with
non-integer values of $\alpha_1$ and $\alpha_2$, 
we have the fractional screening function (\ref{ff2}).

As a result, the exponential functions can be used in
the generalized polarizable point dipoles method for  
molecular dynamics and Monte Carlo simulations 
of low-loss dielectric materials.

\newpage


\begin{thebibliography}{**}


\bibitem{W1} P. Debye, 
"Some results of kinetic theory of isolators. (preliminary announcement)" 
Physikalische Zeitschrift {\bf 13} (1912) 97-100; \\
P. Debye, E. Huckel,
"The theory of electrolytes I. 
The lowering of the freezing point and related occurrences"
Physikalische Zeitschrift  {\bf 24} (1923) 185-206. 

\bibitem{Jo2} A.K. Jonscher, 
{\it Universal Relaxation Law}, 
(Chelsea Dielectrics Press, London, 1996).

\bibitem{Jrev} A.K. Jonscher, 
"Dielectric relaxation in solids" 
J. Physics D  Appl. Phys. {\bf 32} (1999) R57-R70. 

\bibitem{Ram} T.V. Ramakrishnan, M.R. Lakshmi, (Eds.), 
{\it Non-Debye Relaxation in Condensed Matter}
(World Scientific, Singapore, 1984).

\bibitem{Jo4} A.K. Jonscher, 
"Universal dielectric response" 
Nature {\bf 267} (1977) 673-679; 
"Low-frequency dispersion in carrier-dominated dielectrics" 
Philosophical Magazine B {\bf 38} (1978) 587-601.

\bibitem{Jo6}  K.L. Ngai, A.K. Jonscher, C.T. White,
"Origin of the universal dielectric response in condensed matter"
Nature {\bf 277} (1979) 185-189. 


\bibitem{Bergman} R. Bergman,
"General susceptibility functions for relaxations in disordered systems"
J Appl. Phys.  {\bf 88} (2000) 1356-1365.

\bibitem{BSC} F. Brouers, O. Sotolongo-Costa, 
"Relaxation in heterogeneous systems: A rare events dominated phenomenon"
Physica A {\bf 356} (2005) 359-374.

\bibitem{RF2} Ya.E. Ryabov, Yu. Feldman,
"Novel approach to the analysis ofthe non-Debye 
dielectric spectrum broadening"
Physica A {\bf 314} (2002) 370-378.

\bibitem{RF1} Ya.E. Ryabov, Yu. Feldman, 
"The Relationship between  the scaling parameter 
and relaxation time for non-exponential relaxation in disordered systems"
Fractals {\bf 11} (2003) 173-183. 

\bibitem{Frenning} G. Frenning, 
"Dielectric-response function determined by regular singular-point analysis"
Phys. Rev. B {\bf 65} (2002) 245117.


\bibitem{SKM} S.G. Samko, A.A. Kilbas, O.I. Marichev,
{\it Fractional Integrals and Derivatives Theory and Applications}
(Gordon and Breach, New York, 1993).

\bibitem{KST} A.A. Kilbas, H.M. Srivastava, J.J. Trujillo,
{\it Theory and Application of Fractional Differential Equations}
(Elsevier, Amsterdam, 2006).

\bibitem{Zaslavsky1} G.M. Zaslavsky, 
"Chaos, fractional kinetics, and anomalous transport"
Phys. Rep. {\bf 371} (2002) 461-580.

\bibitem{Zaslavsky2}  G.M. Zaslavsky, 
{\it Hamiltonian Chaos and Fractional Dynamics} 
(Oxford University Press, Oxford, 2005).

\bibitem{Mainardi} A. Carpinteri, F. Mainardi, (Eds.)
{\it Fractals and Fractional Calculus in Continuum Mechanics} 
(Springer, New York, 1997).

\bibitem{MK} R. Metzler, J. Klafter, 
"The random walk's guide to anomalous diffusion: 
a fractional dynamics approach" Phys. Rep. {\bf 339} (2000) 1-77;
"The restaurant at the end of the random walk: recent developments 
in the description of anomalous transport by fractional dynamics" 
J. Phys. A {\bf 37} (2004) R161-R208.

\bibitem{Hilfer} {\it Applications of Fractional Calculus in Physics} 
Ed. by R. Hilfer (World Scientific, Singapore, 2000)

\bibitem{Sing} V.E. Tarasov,
"Fractional Calculus and Physics on Fractals"
in "Dynamical Chaos and Non-equilibrium Statistical Mechanics:
From Rigorous Results to Applications in Nano-systems" 
Lecture Notes Series.
(Singapore University Press and World Scientific, Singapore, 2009) 
to be published.

\bibitem{E1} N. Engheta,  
"Fractional curl operator in electromagnetics"
Microwave and Optical Technology Letters
{\bf 17} (1998) 86-91; \\
Q.A. Naqvi, M. Abbas, 
"Complex and higher order fractional curl operator in electromagnetics"
Optics Communications {\bf 241} (2004) 349-355; \\
A. Hussain,  Q.A. Naqvi,
"Fractional curl operator in chiral medium and 
fractional non-symmetric transmission line"
Progress In Electromagnetics Research {\bf 59} (2006) 199-213.

\bibitem{Plasma2005} V.E. Tarasov,
"Electromagnetic field of fractal distribution of charged particles"
Physics of Plasmas {\bf 12} (2005) 082106; 
"Multipole moments of fractal distribution of charges"
Mod. Phys. Lett. B. {\bf 19} (2005) 1107-1118; 
"Magnetohydrodynamics of fractal media"
Physics of Plasmas {\bf 13} (2006) 052107; 
"Electromagnetic fields on fractals"
Mod. Phys. Lett. A. {\bf 21} (2006) 1587-1600.


\bibitem{NR} R.R. Nigmatullin, Ya.E. Ryabov, 
"Cole-Davidson dielectric relaxation as 
a self-similar relaxation process"
Phys. Solid State {\bf 39} (1997) 87-90;
Fiz. Tverd. Tela  {\bf 39} (1997) 101-105 in Russian. 

\bibitem{NP}
V.V. Novikov, V.P. Privalko, 
"Temporal fractal model for the anomalous dielectric relaxation of 
inhomogeneous media with chaotic structure"
Phys. Rev. E. {\bf 64} (2001) 031504.

\bibitem{Dielectric}
Y. Yilmaz, A. Gelir, F. Salehli, R.R. Nigmatullin, A. A. Arbuzov,
"Dielectric study of neutral and charged hydrogels during the swelling process"
J. Chem. Phys.  {\bf 125} (2006) 234705.

\bibitem{First}
R.R. Nigmatullin, A.A. Arbuzov, F. Salehli, A. Giz, I. Bayrak, H. Catalgil-Giz, 
"The first experimental confirmation of the fractional kinetics containing
the complex-power-law exponents: Dielectric measurements of polymerization reactions"
Physica B {\bf 388} (2007) 418-434.


\bibitem{MGo} F. Mainardi, R. Gorenflo, 
"On Mittag-Leffler-type functions in fractional evolution processes" 
J. Comput. Appl. Math. {\bf 118} (2000) 283-299; \\
R. Gorenflo, Y. Luchko, F. Mainardi,
"Wright functions as scale-invariant solutions of the diffusion-wave equation"
J. Computat. Appl. Math. {\bf 118} (2000) 175-191.

\bibitem{TZ2}  V.E. Tarasov, G.M. Zaslavsky,
"Dynamics with low-level fractionality" \
Physica A {\bf 368} (2006) 399-415.

\bibitem{Numer} R. Gorenflo,
"Fractional calculus: some numerical methods"
in A. Carpinteri, F. Mainardi, (Eds.), 
{\it Fractals and Fractional Calculus in Continuum Mechanics}, 
(Springer, Wien and New York, 1997) pp.277-290; \\
O.P. Agrawal, 
"Solution for a fractional diffusion-wave equation defined in a bounded domain" 
Nonlinear Dynamics {\bf 29} (2002) 145-155; \\
C. Tadjeran, M.M. Meerschaert, H.P. Scheffler,
"A second-order accurate numerical approximation for 
the fractional diffusion equation"
Journal of Computational Physics {\bf 213} (2006) 205-213.



\bibitem{Scr1} A.K. Jonscher,  
"Dielectric relaxation with dipolar screening"
J. Material Sci. {\bf 32} (1997) 6409-6414.

\bibitem{Scr2} A.K. Jonscher,  "Low-loss dielectrics"
J. Material Sci. {\bf 34} (1999) 3071-3082.

\bibitem{Mazur} M. Mandel, P. Mazur,  
"On the molecular theory of dielectric polarization"
Physica  {\bf 24} (1958) 116-128.

\bibitem{Ramshaw} J.D. Ramshaw, 
"On the molecular theory of dielectric polarization in rigid-dipole fluids"
J. Chem. Phys. {\bf 55} (1971) 1763-1774.

\bibitem{Applequist} J. Applequist, J.R. Carl, K.K. Fung, 
"An atom dipole interaction model for molecular polarizability. 
Application to polyatomic molecules and determination of atom polarizabilities"
J. Amer. Chem. Soc. {\bf 94} (1972) 2952-2960.

\bibitem{Rick}  S.W. Rick, S.J. Stuart, 
"Potentials and algorithms for incorporating polarizability in computer simulations" 
Rev. Comput. Chem. {\bf 18} (2002) 89-146.

\bibitem{40} J. Applequist, 
"An atom dipole interaction model for molecular optical properties"
Acc. Chem. Res. {\bf 10} (1977) 79-85.

\bibitem{41} B.T. Thole, 
"Molecular polarizabilities calculated with a modified dipole interaction"
Chem. Phys. {\bf 59} (1981) 341-350.

\bibitem{43} F.H. Stillinger, 
"Dynamics and ensemble averages for 
the polarization models of molecular interactions"
J. Chem. Phys. {\bf 71} (1979) 1647-1651. 

\bibitem{G1}  J.C. Burnham, J. Li, S.S. Xantheas, M. Leslie, 
"The parametrization of a Thole-type all-atom polarizable water model from 
first principles and its application to the study of water clusters (n=2–21) 
and the phonon spectrum of ice Ih"
J. Chem. Phys. {\bf 110} (1999) 4566-4581.

\bibitem{G2} P. Ren, J.W. Ponder, 
"A consistent treatment of intra- and intermolecular 
polarization in molecular mechanics calculations"
J. Comput. Chem. {\bf 23} (2002) 1497-1506.

\bibitem{Masia2} M. Masia, M. Probst, R. Rey,
"On the performance of molecular polarization methods.
II. Water and carbon tetrachloride close to a cation"
J. Chem. Phys. {\bf 123} (2005) 164505.

\bibitem{JPA} V.E. Tarasov,
"Continuous limit of discrete systems with long-range interaction" 
J. Phys. A {\bf 39} (2006) 14895-14910. 

\bibitem{Kittel} C. Kittel, {\it Introduction to Solid State Physics}, 
8th edn. (Wiley, New York, 2004).


\end{thebibliography}
\end{document}